\documentclass{IEEEtran}

\usepackage{enumitem,kantlipsum}
\usepackage{graphicx}
\usepackage{epstopdf}
\usepackage{amssymb}
\usepackage{amsmath}
\usepackage{subfigure}
\usepackage{epsfig}
\usepackage{color}
\usepackage{enumitem}
\usepackage{float}
\usepackage{soul}
\usepackage{wrapfig}
\usepackage{arydshln}
\usepackage{tablefootnote}

\usepackage{amsthm}

\def\0{{\bf 0}}
\def\1{{\bf 1}}

\def\beq{\begin{equation*}}
    \def\eeq{\end{equation*}}
\def\bql{\begin{equation}}
    \def\eql{\end{equation}}
\def\bqn{\begin{eqnarray*}}
    \def\eqn{\end{eqnarray*}}
\def\bnl{\begin{eqnarray}}
    \def\enl{\end{eqnarray}}
\def\bma{\begin{bmatrix}}
    \def\ema{\end{bmatrix}}
\def\bmx{\begin{matrix}}
    \def\emx{\end{matrix}}
\def\ben{\begin{enumerate}}
    \def\een{\end{enumerate}}
\def\bit{\begin{itemize}}
    \def\eit{\end{itemize}}
\def\bei{\begin{itemize}}
    \def\eei{\end{itemize}}
\def\bet{\begin{tabular}}
    \def\eet{\end{tabular}}

\newcommand{\ba}{\mathbf{a}}

\newcommand{\be}{\mathbf{e}}

\usepackage[dvipsnames]{xcolor}

\def\1{{\bf1}}

\def\b{{\beta}}
\def\a{\alpha}

\def\bit{\begin{itemize}}
\def\eit{\end{itemize}}

\def\be{\begin{equation}}
\def\ee{\end{equation}}
\def\ba{\begin{eqnarray}}
\def\ea{\end{eqnarray}}

\def\bes{\begin{equation*}}
\def\ees{\end{equation*}}
\def\bas{\begin{eqnarray*}}
\def\eas{\end{eqnarray*}}

\usepackage{url}
\usepackage{verbatim}

\newtheorem{Remark 1}{Remark}
\newtheorem{Remark 2}[Remark 1]{Remark}
\newtheorem{Remark 3}[Remark 1]{Remark}
\newtheorem{Remark 4}[Remark 1]{Remark}
\newtheorem{Remark 5}[Remark 1]{Remark}
\newtheorem{Remark 6}[Remark 1]{Remark}
\newtheorem{Remark 7}[Remark 1]{Remark}

\newtheorem{Lemma 1}{Lemma}
\newtheorem{Lemma 2}[Lemma 1]{Lemma}
\newtheorem{Lemma 3}[Lemma 1]{Lemma}
\newtheorem{Lemma 4}[Lemma 1]{Lemma}
\newtheorem{Lemma 5}[Lemma 1]{Lemma}
\newtheorem{Lemma 6}[Lemma 1]{Lemma}
\newtheorem{Lemma 7}[Lemma 1]{Lemma}

\newtheorem{Assumption 1}{Assumption}
\newtheorem{Assumption 2}[Assumption 1]{Assumption}
\newtheorem{Assumption 3}[Assumption 1]{Assumption}
\newtheorem{Assumption 4}[Assumption 1]{Assumption}
\newtheorem{Definition 1}{Definition}
\newtheorem{Theorem 1}{Theorem}
\newtheorem{Theorem 2}[Theorem 1]{Theorem}
\newtheorem{Theorem 3}[Theorem 1]{Theorem}
\newtheorem{Theorem 4}[Theorem 1]{Theorem}
\newtheorem{Theorem 5}[Theorem 1]{Theorem}
\newtheorem{Theorem 6}[Theorem 1]{Theorem}
\newtheorem{Theorem 7}[Theorem 1]{Theorem}
\newtheorem{Theorem 8}[Theorem 1]{Theorem}
\newtheorem{Theorem 9}[Theorem 1]{Theorem}
\newtheorem{Theorem 10}[Theorem 1]{Theorem}

\newtheorem{Proposition 1}{Proposition}

\title{\LARGE \bf
 A Robust Dynamic Average Consensus Algorithm that  Ensures
both Differential Privacy and Accurate Convergence}

\author{Yongqiang Wang
\thanks{An extended version of the work (with additional applications to Generalized Nash Equilibrium seeking) has been submitted to IEEE Transactions on Automatic Control \cite{wang2022ensure}. \\

\vspace{-0.3cm}
 The work   was supported in part by the National Science Foundation under Grants   ECCS-1912702, CCF-2106293, CCF-2215088, and
CNS-2219487.}
\thanks{Yongqiang Wang is with the Department of Electrical and Computer Engineering, Clemson University, Clemson, SC 29634, USA
{\tt\small{yongqiw}@clemson.edu}
}%

}

\begin{document}

\maketitle
\thispagestyle{empty}
\pagestyle{empty}

\begin{abstract}
We propose a new dynamic average consensus algorithm that is robust to information-sharing noise arising from  differential-privacy design. Not only is dynamic average consensus widely used in cooperative control and distributed tracking, it is also a   fundamental building block in numerous distributed computation algorithms such as multi-agent optimization and distributed Nash equilibrium seeking.  We propose a new dynamic average consensus algorithm that is robust to persistent and independent information-sharing noise added for the purpose of differential-privacy protection. In fact, the algorithm  can ensure both provable convergence to the exact average reference signal and rigorous $\epsilon$-differential privacy (even when the number of iterations tends to infinity), which, to our knowledge,   has not been achieved before in average consensus algorithms.  Given that channel noise in communication can  be viewed  as a special case of differential-privacy noise, the algorithm can also be used to   counteract communication imperfections.  Numerical simulation results   confirm the effectiveness of the proposed approach.

\end{abstract}

\section{Introduction}

 Dynamic average consensus addresses the problem where a group of agents cooperate to track
the average of multiple time-varying reference signals while each individual agent can only access one reference signal and  communicate with its immediate neighbors \cite{zhu2010discrete,kia2019tutorial}. Compared with the intensively studied static average consensus problem where the reference signals are confined to be the time-invariant initial values of individual agents, the dynamic average consensus problem allows the tracking targets to evolve with time and fits naturally into   applications  such as formation control \cite{yang2007distributed}, sensor fusion \cite{olfati2005consensus}, and distributed tracking \cite{george2019robust}. More interestingly, in recent years,   dynamic average consensus is also proven to be an effective primitive in constructing fully distributed computation algorithms for networked games \cite{koshal2016distributed,belgioioso2020distributed} and large-scale optimization \cite{xu2015augmented,di2016next,pu2020push}.

The vast applicability of dynamic average consensus has spurred intensive interests, and plenty of results have been proposed (see, e.g., the survey paper \cite{kia2019tutorial}). However, all of these results require participating agents to exchange and disclose their states
explicitly to neighboring agents in every iteration,  which is problematic when involved information is sensitive. For example, in  sensor network based source localization, disclosing intermediate states will make individual agents' positions inferrable, which is undesirable
when sensors want to keep their positions private in sensitive applications \cite{zhang2019admm,burbano2019inferring}. Another example is smart grids where
multiple power generators have to reach agreement on cost while preserving the privacy of their individual generation information, which is sensitive in bidding the right for energy selling \cite{fang2011smart}. The privacy problem is more pronounced in distributed optimization and learning (many of which are built upon dynamic average consensus) in that the shared states in distributed optimization and learning  may contain sensitive information such as  salary information or medical records \cite{yan2012distributed}. In fact, it has been shown in \cite{wang2022tailoring,wang2022decentralized} that without a privacy mechanism in place,
an adversary  (e.g., the DLG attacker in \cite{zhu2019deep})  can use shared state information to precisely recover the   raw data used for training.

To address the privacy issue in dynamic average consensus, several approaches have been proposed  (see, e.g., \cite{kia2015dynamic,zhang2022privacy}). In fact, given that the information-sharing mechanism in dynamic average consensus is the  same as in static average consensus, many privacy solutions designed for static average consensus (e.g., our prior work \cite{gao2018privacy,ruan2019secure,wang2019privacy,gao2022algorithm} as well as others' \cite{manitara2013privacy,pequito2014design,mo2016privacy,nozari2017differentially,gupta2017privacy,altafini2019dynamical,he2018privacy}) can also be applied directly to dynamic average consensus.  However, these privacy approaches are restricted  in that they either require the communication graph to satisfy certain properties, or can only protect the exact sensitive (initial) value  from being {\it uniquely} inferrable by the adversary.   As differential privacy (DP) has become the de facto standard for privacy protection due to its  strong resilience against arbitrary post-processing and  auxiliary information \cite{dwork2014algorithmic}, plenty of results have emerged on differentially-private average consensus (see, e.g. \cite{nozari2017differentially,huang2012differentially,fiore2019resilient,he2020differential,liu2020differentially}).
    However, all of these results have to sacrifice provable convergence to the exact consensus value to enable rigorous DP (to ensure a finite privacy budget when the number of iterations tends to infinity). Moreover,   these results only address the static average consensus problem, where the information to be protected are time-invariant initial values of individual agents' states.


In this paper, we propose a new dynamic average consensus algorithm that is robust to persistent and independent information-sharing noise injected for the purpose of differential-privacy protection. In fact, the algorithm is proven able to ensure both provable convergence to the desired average reference signal
and rigorous $\epsilon$-DP, with the cumulative privacy budget guaranteed to be finite even when the number of iterations tends to infinity.  The approach is motivated by the observation that DP-noise  enters the algorithm
through inter-agent interaction, and hence, its influence on convergence accuracy can be attenuated
by gradually weakening inter-agent interaction. Not  only is the proposed algorithm  the first to achieve rigorous $\epsilon$-DP in dynamic average consensus, it can also retain provable convergence to the {\it exact} average reference signal when  reference signals' variations decay with time. To our knowledge, ensuring accurate consensus  while enabling rigorous $\epsilon$-DP  has not been reported before for dynamic average consensus.


{\bf Notations:}
 We use $\mathbb{R}^d$ to denote the Euclidean space of
dimension $d$. We write $I_d$ for the identity matrix of dimension $d$,
and ${\bf 1}_d$ for  the $d$-dimensional  column vector with all
entries equal to 1; in both cases  we suppress the dimension when it is
clear from the context. 
A vector is viewed as a column
vector, and for  a
vector $x$, $[x]_i$ denotes its $i$th element.
  We use $\langle\cdot,\cdot\rangle$ to denote the inner product and
 $\|x\|$ for the standard Euclidean norm of a vector $x$. We use $\|x\|_1$ to represent the $L_1$ norm of a vector $x$.
We write $\|A\|$ for the matrix norm induced by the vector norm $\|\cdot\|$.
 $A^T$ denotes the transpose of a matrix $A$.
 Given vectors $x_1,\cdots,x_m$, we define $\bar{x}=\frac{\sum_{i=1}^{m}x_i}{m}$.
  Often, we abbreviate {\it almost surely} by {\it a.s}.
\def\as{{\it a.s.\ }}


\section{Problem Formulation and Preliminaries}\label{sec-problem}
\subsection{On Dynamic Average Consensus}

We consider a dynamic average consensus problem  among a set of $m$   agents  $[m]=\{1,\,\cdots,m\}$. We index the agents by $1,\,2,\,\cdots,m$. Agent $i$   can access fixed-frequency samples of its own reference signal $r_i\in\mathbb{R}^d$, which could be  varying with time. Every agent $i$ also maintains a state $x_i\in\mathbb{R}^d$. The aim of dynamic average consensus is for all agents to collaboratively track the average reference signal $\bar{r}\triangleq \frac{\sum_{i=1}^{m} {r_i}}{m}$ while every agent can only access discrete-time measurements of its own reference signal and share its state with its immediate neighboring agents.

  We describe the local communication among agents using a weight matrix
$L=\{L_{ij}\}$, where $L_{ij}>0$ if agent $j$ and agent $i$ can directly communicate with each other,
and $L_{ij}=0$ otherwise. For an agent $i\in[m]$,
its  neighbor set
$\mathbb{N}_i$ is defined as the collection of agents $j$ such that $L_{ij}>0$.
We define $L_{ii}\triangleq-\sum_{j\in\mathbb{N}_i}L_{ij}$  for all $i\in [m]$,
where $\mathbb{N}_i$ is the neighbor set of agent $i$. Furthermore,
We make the following assumption on $L$:

\begin{Assumption 1}\label{as:L}
 The matrix  $L=\{w_{ij}\}\in \mathbb{R}^{m\times m}$ is symmetric and satisfies
    ${\bf 1}^TL={\bf
  0}^T$, $L{\bf 1}={\bf
  0}$, and $ \|I+L-\frac{{\bf 1}{\bf 1}^T}{m}\|<1$.
\end{Assumption 1}

Assumption~\ref{as:L} ensures that the interaction graph induced by $L$ is connected, i.e., there is a  path
from each agent to every other agent.
It can be verified that $\|I+L-\frac{{\bf 1}{\bf 1}^T}{m}\|=\max\{|1+\rho_2|,|1+\rho_m|\}$, where $\{\rho_i,i\in[m]\}$ are the eigenvalues of $L$, with
$\rho_m\le \ldots\le \rho_2<\rho_1=0$.

We also need  the following lemma for convergence analysis:
\begin{Lemma 2}\cite{wang2022tailoring}\label{Lemma-polyak}
Let $\{v^k\}$,$\{o^k\}$, and $\{p^k\}$ be random nonnegative scalar sequences, and
$\{q^k\}$ be a deterministic nonnegative scalar sequence satisfying
$\sum_{k=0}^\infty o^k<\infty$  {\it a.s.},
$\sum_{k=0}^\infty q^k=\infty$, $\sum_{k=0}^\infty p^k<\infty$ {\it a.s.},
and
\[
\mathbb{E}\left[v^{k+1}|\mathcal{F}^k\right]\le(1+o^k-q^k) v^k +p^k,\quad \forall k\geq 0\quad\as
\]
where $\mathcal{F}^k=\{v^\ell,o^\ell,p^\ell; 0\le \ell\le k\}$.
Then, $\sum_{k=0}^{\infty}q^k v^k<\infty$ and
$\lim_{k\to\infty} v^k=0$ hold almost surely.
\end{Lemma 2}

\begin{Lemma 4}\cite{wang2022tailoring}\label{le:chung}
Let $\{v^k\}$ be a nonnegative sequence, and  $\{\a^k\}$  and $\{\b^k\}$ be positive sequences satisfying $\sum_{k=0}^{\infty}\a^k=\infty$, $\lim_{k\rightarrow \infty} \a^k =0$, and $\lim_{k\rightarrow \infty}\frac{\b^k}{\a^k}=0$. If there exists a $K\geq 0$ such that $ v^{k+1} \le(1-\a^k) v^k +\b^k$ holds for all $k\geq K$,
 then we always have $v^k\leq C \frac{\b^k}{\a^k}$  for all $k$, where $C$ is some constant.
\end{Lemma 4}

\subsection{On Differential Privacy}
We consider  adversaries having full access to all  communication channels. Namely, the adversary  can peek inside all the messages going back and forth between the agents. Because in dynamic average consensus, the sensitive information of participating agents are time-varying reference signals,   we use the notion of $\epsilon$-DP for continuous bit streams \cite{dwork2010differential}, which has recently been applied to distributed optimization  (see \cite{Huang15} as well as our own work \cite{wang2022tailoring}).
To enable $\epsilon$-DP, we use the Laplace noise  mechanism, i.e., we add Laplace noise to all shared messages.
For a constant $\nu>0$, we use ${\rm Lap}(\nu)$ to denote a Laplace distribution of a scalar random variable with the probability density function $x\mapsto\frac{1}{2\nu}e^{-\frac{|x|}{\nu}}$. One can verify that ${\rm Lap}(\nu)$'s mean is   zero   and its  variance is $2\nu^2$.
Inspired by the formulation of distributed optimization  in \cite{Huang15}, to facilitate DP analysis, we represent a dynamic average consensus problem $\mathcal{P}$  by  $\mathcal{P}\triangleq (\mathcal{J}, L)$, where
 $ \mathcal{J} \triangleq\{ r_1,\,\cdots,r_m\}$ are the reference signals of all agents and $ L$ is the inter-agent interaction weight matrix. Then, we define adjacency between two dynamic average consensus problems as follows:

\begin{Definition 1}\label{de:adjacency}
Two dynamic average consensus problems $\mathcal{P}=(\mathcal{J}, L)$ and $\mathcal{P}'=(\mathcal{J}', L')$ are adjacent if the following conditions hold:
\begin{itemize}
\item   $L=L'$, i.e., the  interaction weight matrices are identical;
\item there exists an $i\in[m]$ such that $r_i\neq r_i'$ but $r_j=r_j'$ for all $j\in[m],\,j\neq i$;
\item  the different reference signals $r_i$ and $r'_i$  have similar steady-state behaviors.
   More specifically, there exists some positive $C_r$ and non-negative non-summable but square-summable sequences $\{\chi^k\}$ and $\{\gamma^k\}$ (which will be specified later in Algorithm 1 and Assumption \ref{as:reference_signal}, respectively)
   such that $\|r_i^k-{r_i'}^k\|_1\leq C_r\chi^k\gamma^k$  holds for all $k$.
\end{itemize}
\end{Definition 1}

 \begin{Remark 1}
 In Definition \ref{de:adjacency}, since  the change  from $r_i$ to $r'_i$ in the second condition  can be arbitrary,  the third condition is added to restrict the change magnitude. It is necessary to enabling rigorous $\epsilon$-DP while maintaining accurate convergence.  This is because  DP aims to make observations statistically indistinguishable while accurate convergence means that the state will stop changing and remain time-invariant after a transient period. Hence,  to make the observations of $\mathcal{P}$ and $\mathcal{P}'$ the same after their states converge and remain  at  their respective converging points,  we have to require $\mathcal{P}$ and $\mathcal{P}'$ to have identical converging points.
\end{Remark 1}

 Given a dynamic average consensus algorithm, we represent an execution of such an algorithm as $\mathcal{A}$, which is an infinite sequence of the iteration variable $\vartheta$, i.e., $\mathcal{A}=\{\vartheta^0,\vartheta^1,\cdots\}$. We consider adversaries that can observe all communicated messages among the agents. Therefore, the observation part of an execution is the infinite sequence of shared messages, which is denoted as $\mathcal{O}$. Given a dynamic average consensus  problem $\mathcal{P}$ and an initial state $\vartheta^0$, we define the observation mapping as $\mathcal{R}_{\mathcal{P},\vartheta^0}(\mathcal{A})\triangleq \mathcal{O}$. Given a dynamic average consensus  problem $\mathcal{P}$, observation sequence $\mathcal{O}$, and an initial state $\vartheta^0$,  $\mathcal{R}_{\mathcal{P},\vartheta^0}^{-1}(\mathcal{O})$ is the set of executions $\mathcal{A}$ that can generate the observation $\mathcal{O}$.
 \begin{Definition 1}\label{de:differential_privacy}
   ($\epsilon$-differential privacy, adapted from \cite{Huang15}). For a given $\epsilon>0$, an iterative distributed algorithm  is $\epsilon$-differentially private if for any two adjacent $\mathcal{P}$ and $\mathcal{P}'$, any set of observation sequences $\mathcal{O}_s\subseteq\mathbb{O}$ (with $\mathbb{O}$ denoting the set of all possible observation sequences), and any initial state ${\vartheta}^0$, the following relationship always holds
    \begin{equation}
        \mathbb{P}[\mathcal{R}_{\mathcal{P},\vartheta^0}^{-1}\left(\mathcal{O}_s\right)]\leq e^\epsilon\mathbb{P}[\mathcal{R}_{\mathcal{P}',\vartheta^0}^{-1}\left( \mathcal{O}_s\right)],
    \end{equation}
    with the probability $\mathbb{P}$  taken over the randomness over iteration processes.
 \end{Definition 1}

 $\epsilon$-DP ensures that an adversary having access to all shared messages cannot gain information with a  significant probability of any participating agent's reference signal. It can  be seen that a smaller $\epsilon$ means a higher level of privacy protection. It is  worth noting that the considered notion of $\epsilon$-DP is more stringent than other relaxed (approximate) DP notions such as $(\epsilon,\,\delta)$-DP \cite{kairouz2015composition}, zero-concentrated DP \cite{bun2016concentrated}, or R\'{e}nyi DP \cite{mironov2017renyi}.


\section{Differentially-private dynamic average consensus}\label{se:algorithm1}
In this section, we present our new dynamic average consensus algorithm that is robust to DP-noise, which is summarized in Algorithm 1.  Our basic idea is to use a decaying factor $\chi^k$ to  gradually attenuate inter-agent interaction, and hence, attenuate the influence of DP-noise. Of course, to enable necessary information fusion among the agents, we have to judiciously design the decaying factor. Moreover, in our proposed algorithm, another fundamental  difference from  conventional dynamic average consensus algorithms  is the introduction of a stepsize $\alpha^k$. The two  features enable Algorithm 1 to avoid the accumulation  and   explosion of noise variance in  conventional dynamic average consensus algorithms in the presence of noise. In fact, the  proposed new algorithm  can guarantee convergence to the {\it exact} average reference signal even when the  DP-noise is allowed to increase with time to ensure rigorous $\epsilon$-DP, which will be elaborated later in detail in Sec. V.  To our knowledge, this is the first time that both provable convergence and rigorous $\epsilon$-DP are achieved in dynamic average consensus algorithms. 

\noindent\rule{0.49\textwidth}{0.5pt}
\noindent\textbf{Algorithm 1:
Robust dynamic average consensus}
\noindent\rule{0.49\textwidth}{0.5pt}
\begin{enumerate}[wide, labelwidth=!, labelindent=0pt]
    \item[] Parameters: Weakening factor $\chi^k>0$ and stepsize $\alpha^k>0$.
    \item[] Every agent $i$'s  reference signal is $r_i^k$. Every agent $i$ maintains one state variable  $x_i^k$, which is initialized as $x_i^0=r_i^0$.
    \item[] {\bf for  $k=1,2,\ldots$ do}
    \begin{enumerate}
        \item Every agent $j$ adds persistent DP-noise   $\zeta_j^{k}$ 
        to its state
    $x_j^k$,  and then sends the obscured state $x_j^k+\zeta_j^{k}$ to agent
        $i\in\mathbb{N}_j$.
        \item After receiving  $x_j^k+\zeta_j^k$ from all $j\in\mathbb{N}_i$, agent $i$ updates its state  as follows:
        \begin{equation}\label{eq:update_in_Algorithm1}
        \begin{aligned}
             x_i^{k+1} &=(1-\alpha^k)x_i^k+\chi^k\sum_{j\in \mathbb{N}_i} L_{ij}(x_j^k+\zeta_j^k-x_i^k)\\
             &\quad+r_i^{k+1}-(1-\alpha^k)r_i^k.
        \end{aligned}
        \end{equation}
    \end{enumerate}
\end{enumerate}
\vspace{-0.1cm} \rule{0.49\textwidth}{0.5pt}

We make the following assumption on the DP-noise:
\begin{Assumption 1}\label{ass:dp-noise}
For every $i\in[m]$ and every $k$, conditional on $x_i^k$,
the  DP-noise $\zeta_i^k$ satisfies $
\mathbb{E}\left[\zeta_i^k\mid x_i^k\right]=0$ and $\mathbb{E}\left[\|\zeta_i^k\|^2\mid x_i^k\right]=(\sigma_{i}^k)^2$ for all  $k\ge0$, and
\begin{equation}\label{eq:condition_assumption1}
\sum_{k=0}^\infty (\chi^k)^2\, \max_{i\in[m]}(\sigma_{i}^k)^2 <\infty,
\end{equation} where $\{\chi^k\}$ is the weakening factor sequence from Algorithm 1.
The initial random vectors satisfy
$\mathbb{E}\left[\|x_i^0\|^2\right]<\infty$,  $\forall i\in[m]$.
\end{Assumption 1}

\section{Convergence Analysis}
In this section, we prove that when the following assumption holds, the proposed algorithm can  ensure every $x_i^k$ to converge almost surely  to the {\it exact} average reference signal $\bar{r}^k \triangleq\frac{1}{m}\sum_{i=1}^{m} r_i^k$:
\begin{Assumption 3}\label{as:reference_signal}
For every $i\in[m]$, there exist   some nonnegative sequence $\{\gamma^k\}$ and a constant $C$ such that
    \begin{equation}\label{eq:bound_C}
       \|r_i^{k+1}-(1-\alpha^k)r_i^{k}\|\leq \gamma^k C
    \end{equation}
  holds,   where $\alpha^k$ is from Algorithm 1 and $\{\gamma^k\}$  satisfies $\lim_{k\rightarrow\infty}\frac{\alpha^k}{\gamma^k}<\infty$.
\end{Assumption 3}

It can be seen that  the condition $\lim_{k\rightarrow\infty}\frac{\alpha^k}{\gamma^k}<\infty$ is necessary since otherwise (\ref{eq:bound_C}) will not hold for constant $r^k$.

 To prove that the state $x_i^k$   converges to the precise average $\bar{r}^k$, we first  prove that $\bar{x}^k=\frac{\sum_{i=1}^{m}x_i^k}{m}$  converges to   $\bar{r}^k$.
 Such a property holds for the conventional dynamic average consensus in the absence of DP-noise. However, it does not hold any more in the presence of information-sharing  noise since the noise will accumulate on $\bar{x}^k$ in the conventional dynamic average consensus algorithm, leading to an exploding variance  \cite{pu2020robust,wang2022gradient} (see also the blue curves in the numerical simulation results in Fig. \ref{fig:tracking_error}).  Here, we prove that   our proposed algorithm can ensure the converge of $\bar{x}^k$ to $\bar{r}^k$ even when  all agents add independent DP-noise  to their shared messages.

\begin{Lemma 1}\label{le:bar_x=bar_r}
Under Assumptions \ref{as:L}, \ref{ass:dp-noise},  $\bar{x}^k$ in the proposed algorithm converges {\it a.s.} to $\bar{r}^k$ if the following conditions hold:
\[
 \sum_{k=0}^{\infty}\alpha^k= \infty, \quad \sum_{k=0}^{\infty}(\alpha^k)^2<\infty.
\]
\end{Lemma 1}
\begin{proof}

According to the definitions of $\bar{x}^k$ and $\bar{r}^k$, we have the following relationship based on (\ref{eq:update_in_Algorithm1}):
\begin{equation}\label{eq:conservation_average}
\bar{x}^{k+1}=(1-\alpha^k)\bar{x}^k+\chi^k\bar{\zeta}^k+\bar{r}^{k+1}-(1-\alpha^k)\bar{r}^k,
\end{equation}
where $\bar{\zeta}^k \triangleq \frac{\sum_{i=1}^{m} |L_{ii}|\zeta_i^k}{m}$ and we have used the symmetric property of $L$ and the definition   $L_{ii}\triangleq-\sum_{j\in\mathbb{N}_i}L_{ij}$.

The preceding relationship implies
\[
\begin{aligned}
\|\bar{x}^{k+1}-\bar{r}^{k+1}\|^2&=\|(1-\alpha^k)(\bar{x}^k-\bar{r}^k)+\chi^k\bar{\zeta}^k\|^2 \\
&=(1-\alpha^k)^2\|\bar{x}^k-\bar{r}^k\|^2+(\chi^k)^2\|\bar{\zeta}^k\|^2\\
&\qquad+2\left\langle  (1-\alpha^k)(\bar{x}^k-\bar{r}^k),\chi^k\bar{\zeta}^k \right\rangle.
\end{aligned}
\]
Using the assumption that the DP-noise $\zeta_i^k$ has zero mean and variance
$(\sigma_{i}^k)^2$ conditional  on $x_i^k$
(see Assumption~\ref{ass:dp-noise}), taking the conditional expectation, given $\mathcal{F}^k=\{x^0,\,\ldots,x^k\}$,
 we obtain  the following inequality  for all $k\ge0$:
 \[
\begin{aligned}
\mathbb{E}\left[\|\bar{x}^{k+1}-\bar{r}^{k+1}\|^2|\mathcal{F}^k\right]&
 \leq\left(1+(\alpha^k)^2-2\alpha^k\right) \|\bar{x}^k-\bar{r}^k\|^2\\
 &\qquad+\frac{\sum_{i=1}^{m}L_{ii}^2(\sigma_i^k)^2(\chi^k)^2}{m},
\end{aligned}
\]
where we have used the relationship $\|\bar{\zeta}^k\|^2\leq \frac{\sum_{i=1}^{m}L_{ii}^2\| \zeta_i^k\|^2}{m}$.

 Under the conditions in the lemma statement, it can be seen that  $\|\bar{x}^{k+1}-\bar{r}^{k+1}\|^2$ satisfies the conditions for $v^k$ in Lemma \ref{Lemma-polyak}, with $o^k=(\alpha^k)^2$, $q^k=2\alpha^k$, and $p^k=\frac{\sum_{i=1}^{m}L_{ii}^2(\sigma_i^k)^2(\chi^k)^2}{m}$. Therefore, we have $\|\bar{x}^{k+1}-\bar{r}^{k+1}\|^2$ converging {\it a.s.} to zero, and hence, $\bar{x}^k$ converging {\it a.s.} to $\bar{r}^k$.
 \end{proof}

\begin{Lemma 1}\label{Le:rho_2}
Under Assumption \ref{as:L} and two positive sequences $\{\chi^k\}$ and $\{\alpha^k\}$ satisfying $
\sum_{k=0}^\infty (\alpha^k)^2<\infty
$ and $
\sum_{k=0}^\infty (\chi^k)^2<\infty
$, there always exists a $T>0$ such that
$
\|(1-\alpha^k)(I- \frac{{\bf 1}{\bf 1}^T}{m})+\chi^kL\|< 1-\chi^k |\rho_2|
$
holds for all $k\geq T$, where $\rho_2$ is the   second largest eigenvalue of $L$.
\end{Lemma 1}
\begin{proof}
Because $L$ is symmetric according to Assumption \ref{as:L},
all of its eigenvalues  are real numbers.
Since $L$' off-diagonal entries are  non-negative and its  diagonal entries $L_{ii}$ are given by
$L_{ii}=-\sum_{j\in\mathbb{N}_i}L_{ij}$, the Gershgorin circle theorem implies  that  all eigenvalues of $L$ are non-positive, with one of them being equal to 0. In fact, the $0$ eigenvalue  is simple (i.e., all the other eigenvalues are strictly negative) from Assumption \ref{as:L}. Arrange the eigenvalues of $L$ as $\rho_m\leq \rho_{m-1}\leq \cdots\leq \rho_2<\rho_1=0$. It can be verified that the eigenvalues of $(1-\alpha^k)I+\chi^kL$ are equal to $1-\alpha^k+\chi^k\rho_m\leq 1-\alpha^k+\chi^k\rho_{m-1}\leq \cdots\leq 1-\alpha^k+\chi^k\rho_2< 1-\alpha^k+\chi^k\rho_1=1-\alpha^k$, and the eigenvalues of $(1-\alpha^k)(I-\frac{{\bf 1}{\bf 1}^T}{m})+\chi^kL=(1-\alpha^k)I+\chi^kL+(1-\alpha^k) \frac{{\bf 1}{\bf 1}^T}{m} $ are given by $\{1-\alpha^k+\chi^k\rho_m,\,1-\alpha^k+\chi^k\rho_{m-1},\, \cdots,\,1-\alpha^k+\chi^k\rho_2,\, 0\}$, with $|\rho_m|\geq|\rho_{m-1}|\geq\cdots\geq |\rho_2|>0$. Hence, the norm of $\|(1-\alpha^k)(I-\frac{{\bf 1}{\bf 1}^T}{m})+\chi^kL\|$ is no larger than $\max\{|1-\alpha^k+\chi^k\rho_m|, |1-\alpha^k+\chi^k\rho_2|\}$. Further taking into account the fact that $\alpha^k$ and $\chi^k$ decay  to zero because they are square summable, we conclude that there always exists a $T>0$ such that $|1-\alpha^k+\chi^k \rho_m|=1-\alpha^k-\chi^k|\rho_m|<1- \chi^k|\rho_m|$ and $|1-\alpha^k+\chi^k \rho_2|=1-\alpha^k-\chi^k|\rho_2|<1- \chi^k|\rho_2|$ hold   for all $k\geq T$. Given $|\rho_m|\geq  |\rho_2|$,   we have the stated result.
\end{proof}

\begin{Theorem 1}\label{Th:consensus_tracking}
 Under Assumptions \ref{as:L}, \ref{ass:dp-noise}, \ref{as:reference_signal}, if
   the nonnegative sequences $\{\alpha^k\}$  and $\{\chi^k\}$ in Algorithm 1 and the nonnegative sequence $\{\gamma^k\}$ in Assumption \ref{as:reference_signal} satisfy
    \begin{equation}\label{eq:condtions_chi}
    \sum_{k=0}^{\infty}\alpha^k=\infty,\, \sum_{k=0}^{\infty}\chi^k=\infty,\, \sum_{k=0}^{\infty}(\chi^k)^2<\infty,\, \sum_{k=0}^{\infty}\frac{(\gamma^k)^2}{\chi^k}<\infty,
    \end{equation}
   and $\lim_{k\rightarrow\infty}\frac{\alpha^k}{\gamma^k}<\infty$, then, the following results hold \as:
     \begin{enumerate}
    \item  every  $x_i^k$ in Algorithm 1 converges to $\bar{r}^k=\frac{\sum_{i=1}^{m}r_i^k}{m}$;
    \item    $\sum_{k=0}^{\infty}\chi^k\sum_{i=1}^{m}\|x_i^k-\bar{x}^k\|^2<\infty$;
    \item   $\sum_{k=0}^{\infty}\gamma^k \sum_{i=1}^{m}\|x_i^k-\bar{x}^k\| <\infty$.
  \end{enumerate}
\end{Theorem 1}

\begin{proof}
For the convenience of analysis, we write the iterates of  $x_i^k$ on per-coordinate expressions. More specifically, for all
$\ell=1,\ldots,d,$ and $k\ge0$, we define
$
x^k(\ell)=\left[[x_1^k]_\ell,\ldots,[x_m^k]_\ell\right]^T
$
where $[x_i^k]_\ell$ represents the $\ell$th element of the vector $x_i^k$. Similarly, we  define
$r^k(\ell)=\left[[r_1^k]_\ell,\ldots,[r_m^k]_\ell\right]^T$  and $\zeta^k(\ell) =\left[[\zeta_1^k]_\ell,\ldots,[\zeta_m^k]_\ell\right]^T$. In
this per-coordinate view, (\ref{eq:update_in_Algorithm1})  has the following form  for all $\ell=1,\ldots,d,$
and $k\ge0$:
\begin{equation}\label{eq:update_percord}
\begin{aligned}
x^{k+1}(\ell)=&(1-\alpha^k)x^k(\ell)+\chi^k L x^k(\ell)+\chi^kL^0 \zeta^k(\ell)  \\
&\quad+r^{k+1}(\ell)-(1-\alpha^k)r^k(\ell).
\end{aligned}
\end{equation}
where $L^0$ is the same as $L$ except that all of its diagonal entries are zero. 

The dynamics of   $\bar{x}^k$ is governed by
\begin{equation}\label{eq:bar_x}
\begin{aligned}
[\bar{x}^{k+1}]_\ell=&\frac{{\bf 1}^T}{m} x^{k+1}(\ell)=\frac{{\bf 1}^T}{m}\left((1-\alpha^k)x^k(\ell)+\chi^k L x^k(\ell) \right)\\
&+\frac{{\bf 1}^T}{m}\left(\chi^kL^0 \zeta^k(\ell) + r^{k+1}(\ell)-(1-\alpha^k)r^k(\ell) \right),
\end{aligned}
\end{equation}
where $[\bar{x}^{k+1}]_\ell$ represents the $\ell$-th element of $\bar{x}^{k+1}$.

  Assumption \ref{as:L} implies ${\bf 1}^TL=0$, which can further simplify (\ref{eq:bar_x}) as:
\begin{equation}\label{eq:bar_x2}
\begin{aligned}
 [\bar{x}^{k+1}]_\ell=& \frac{{\bf 1}^T}{m}\left((1-\alpha^k)x^k(\ell)+ \chi^kL^0 \zeta^k(\ell)\right.\\
 &\qquad\left.+  r^{k+1}(\ell)-(1-\alpha^k) r^k(\ell) \right).
\end{aligned}
\end{equation}

 Combining the relations in (\ref{eq:update_percord})  and (\ref{eq:bar_x2}) yields
\begin{equation}\label{eq:v_k+1}
\begin{aligned}
&x^{k+1}(\ell)-{\bf 1}[\bar{x}^{k+1}]_{\ell}\\
&= \left((1-\alpha^k)\Pi+\chi^kL\right)x^k(\ell)+\chi^k \Pi L^0\zeta^k(\ell)\\
&\quad+\Pi\left(r^{k+1}(\ell)
- (1-\alpha^k)r^k(\ell)\right).
\end{aligned}
\end{equation}
where $\Pi\triangleq I-\frac{{\bf 1}{\bf 1}^T}{m}$.

To simplify the  notation, we define
\begin{equation}\label{eq:W^k}
W^k\triangleq (1-\alpha^k)\Pi+\chi^kL=(1-\alpha^k)(I-\frac{{\bf 1}{\bf 1}^T}{m})+\chi^kL.
\end{equation}
Assumption \ref{as:L} ensures $ W^k {\bf 1}=0$, and hence $ W^k {\bf 1}[\bar{x}^k]_\ell=0$  for any $1\leq \ell\leq d$. Subtracting $ W^k {\bf 1}[\bar{x}^k]_\ell=0$  from the right hand side of (\ref{eq:v_k+1}) yields
\begin{equation}\label{eq:v_k+1_3}
\begin{aligned}
x^{k+1}(\ell)-{\bf 1}[\bar{x}^{k+1}]_{\ell}=&W^k\left(x^k(\ell)-{\bf 1}[\bar{x}^k]_\ell\right)+\chi^k \Pi L^0\zeta^k(\ell)\\
& +\Pi \left(r^{k+1}(\ell)
- (1-\alpha^k)r^k(\ell)\right),
\end{aligned}
\end{equation}
which, further implies
\begin{equation}\label{eq:v_k+1_norm}
\begin{aligned}
&\|x^{k+1}(\ell)-{\bf 1}[\bar{x}^{k+1}]_{\ell} \|^2\\
&=\left\|W^k (x^k(\ell)-{\bf 1}[\bar{x}^k]_\ell )+\Pi  (r^{k+1}(\ell)
- (1-\alpha^k)r^k(\ell) ) \right\|^2\\
& +2\left\langle W^k (x^k(\ell)-{\bf 1}[\bar{x}^k]_\ell )+\Pi  (r^{k+1}(\ell)
- (1-\alpha^k)r^k(\ell) ),\right.\\
&\left. \qquad \chi^k \Pi L^0\zeta^k(\ell) \right\rangle+\|\chi^k \Pi L^0\zeta^k(\ell) \|^2\\
&\leq\left\|W^k (x^k(\ell)-{\bf 1}[\bar{x}^k]_\ell )+\Pi  (r^{k+1}(\ell)
- (1-\alpha^k)r^k(\ell) ) \right\|^2\\
& +2\left\langle W^k (x^k(\ell)-{\bf 1}[\bar{x}^k]_\ell )+\Pi  (r^{k+1}(\ell)
-  (1-\alpha^k)r^k(\ell) ),\right.\\
&\left. \qquad \chi^k \Pi L^0\zeta^k(\ell) \right\rangle+(\chi^k)^2\| L^0\|^2\zeta^k(\ell) \|^2.
\end{aligned}
\end{equation}
where we used  $\|\Pi\|=1$.
 Using the assumption that the DP-noise $\zeta_i^k$ has zero mean and variance
$(\sigma_{i}^k)^2$ conditional  on $x_i^k$
(see Assumption~\ref{ass:dp-noise}), taking the conditional expectation, given $\mathcal{F}^k=\{x^0,\,\ldots,x^k\}$,
 we obtain  the following inequality  for all $k\ge0$:
\begin{equation}\label{eq:Ev_k}
\begin{aligned}
&\mathbb{E}\left[\|x^{k+1}(\ell)-{\bf 1}[\bar{x}^{k+1}]_{\ell} \|^2|\mathcal{F}^k\right]\\
&\leq\left\|W^k (x^k(\ell)-{\bf 1}[\bar{x}^k]_\ell )+\Pi  (r^{k+1}(\ell)
- (1-\alpha^k)r^k(\ell) ) \right\|^2\\
&\quad+(\chi^k)^2\|L^0\|^2 \mathbb{E}\left[\|\zeta^k(\ell) \|^2\right] \\
&\leq\left(\|W^k \|\| x^k(\ell)-{\bf 1}[\bar{x}^k]_\ell  \|+ \|   r^{k+1}(\ell)
- (1-\alpha^k)r^k(\ell)  \|\right)^2\\
&\quad+(\chi^k)^2\|L^0\|^2 \mathbb{E}\left[\|\zeta^k(\ell) \|^2\right].
\end{aligned}
\end{equation}

For the first term on the right hand side of the preceding inequality, we bound it using the fact that there exists a $T\geq 0$ such that $0<\|W^k\|\leq 1-\chi^k |\rho_2|$ holds for all $k\geq T$ (see Lemma \ref{Le:rho_2}).  Hence,  equation (\ref{eq:Ev_k}) implies   the following relationship   for all $k\geq T$:
\begin{equation}\label{eq:Ev_k2}
\begin{aligned}
&\mathbb{E}\left[\|x^{k+1}(\ell)-{\bf 1}[\bar{x}^{k+1}]_{\ell} \|^2|\mathcal{F}^k\right]\leq \\
&\left((1\hspace{-0.05cm}-\hspace{-0.05cm}\chi^k|\rho_2|)\|x^k(\ell)\hspace{-0.05cm}-\hspace{-0.05cm}{\bf 1}[\bar{x}^k]_\ell \|+\|r^{k+1}(\ell)
\hspace{-0.05cm}-\hspace{-0.05cm} (1\hspace{-0.05cm}-\hspace{-0.05cm}\alpha^k)r^k(\ell) \|\right)^2\\
&\quad+(\chi^k)^2\|L^0\|^2\mathbb{E}\left[\|\zeta^k(\ell) \|^2\right].
\end{aligned}
\end{equation}

Next, we apply  the inequality
$(a+b)^2\le (1+\epsilon) a^2 + (1+\epsilon^{-1})b^2$, which is valid for any scalars $a,b,$ and $\epsilon>0$, to (\ref{eq:Ev_k2}):
\begin{equation}\label{eq:Ev_k3}
\begin{aligned}
&\mathbb{E}\left[\|x^{k+1}(\ell)-{\bf 1}[\bar{x}^{k+1}]_{\ell} \|^2|\mathcal{F}^k\right]\leq (\chi^k)^2\|L^0\|^2 \mathbb{E}\left[\|\zeta^k(\ell) \|^2\right]\\
&+(1+\epsilon)(1-\chi^k|\rho_2|)^2\|x^k(\ell)-{\bf 1}[\bar{x}^k]_\ell\|^2\\
&+(1+\epsilon^{-1})\|\  r^{k+1}(\ell)
- (1-\alpha^k)r^k(\ell) \|^2.
\end{aligned}
\end{equation}

Setting $\epsilon=\frac{\chi^k|\rho_2|}{1-\chi^k|\rho_2|}$ (which leads to $(1+\epsilon)=\frac{1}{1-\chi^k|\rho_2|}$ and $1+\epsilon^{-1}=\frac{1}{\chi^k|\rho_2|}$) yields
\begin{equation}\label{eq:Ev_k4}
\begin{aligned}
&\mathbb{E}\left[\|x^{k+1}(\ell)-{\bf 1}[\bar{x}^{k+1}]_{\ell} \|^2|\mathcal{F}^k\right]\leq (\chi^k)^2\|L^0\|^2\mathbb{E}\left[\|\zeta^k(\ell) \|^2\right]\\
&+ (1-\chi^k|\rho_2|) \|x^k(\ell)-{\bf 1}[\bar{x}^k]_\ell\|^2\\
&+ \frac{1}{\chi^k|\rho_2|} \| r^{k+1}(\ell)
- (1-\alpha^k)r^k(\ell)  \|^2.
\end{aligned}
\end{equation}

Note that the following relations always hold:
$
\sum_{\ell=1}^d \|x^k(\ell) - [\bar x^k]_\ell{\bf
1}\|^2=\sum_{i=1}^m\|x^k_i - \bar x^k \|^2,$ $
\sum_{\ell=1}^d \|r^{k+1}(\ell) - (1-\alpha^k)r^{k}(\ell)\|^2=\sum_{i=1}^m\|r^{k+1}_i - (1-\alpha^k)r^k_i\|^2$, $\sum_{\ell=1}^d
\|\zeta^k(\ell)\|^2=\sum_{i=1}^m\|\zeta^k_i\|^2$.

Hence,  summing (\ref{eq:Ev_k4}) over $\ell=1,\ldots,d$ leads to
\begin{equation}\label{eq:Ev_k5}
\begin{aligned}
&\mathbb{E}\left[\sum_{i=1}^{m}\|x_i^{k+1}- \bar{x}^{k+1} \|^2|\mathcal{F}^k\right]\\
&\leq (\chi^k)^2\|L^0\|^2\mathbb{E}\left[\sum_{i=1}^{m}\|\zeta_i^k  \|^2\right]+ (1-\chi^k|\rho_2|) \sum_{i=1}^{m}\|x_i^k- \bar{x}^k\|^2\\
&+ \frac{1}{\chi^k|\rho_2|} \sum_{i=1}^{m}\| r_i^{k+1}
- (1-\alpha^k)r_i^k \|^2  \\
&\leq (\chi^k)^2\|L^0\|^2 \sum_{i=1}^{m}(\sigma_i^k)^2+ (1-\chi^k|\rho_2|) \sum_{i=1}^{m}\|x_i^k- \bar{x}^k\|^2\\
&+ \frac{1}{\chi^k|\rho_2|} \sum_{i=1}^{m}\| r_i^{k+1}
- (1-\alpha^k)r_i^k \|^2  .
\end{aligned}
\end{equation}

Using the assumption $\|r_i^{k+1}-(1-\alpha^k)r_i^{k}\|\leq \gamma^k C$ for all $i\in[m]$ from Assumption \ref{as:reference_signal}, we  obtain
\[
\frac{1}{\chi^k|\rho_2|} \sum_{i=1}^{m}\| r_i^{k+1}
- (1-\alpha^k)r_i^k \|^2\leq  \frac{(\gamma^k)^2}{\chi^k|\rho_2|}mC^2.
\]

 Submitting the preceding relationship into (\ref{eq:Ev_k5}) yields
\begin{equation}\label{eq:Ev_k6}
\begin{aligned}
&\mathbb{E}\left[\sum_{i=1}^{m}\|x_i^{k+1}- \bar{x}^{k+1} \|^2|\mathcal{F}^k\right]\leq (\chi^k)^2\|L^0\|^2 \sum_{i=1}^{m}(\sigma_i^k)^2\\ &+(1-\chi^k|\rho_2|) \sum_{i=1}^{m}\|x_i^k- \bar{x}^k\|^2  + \frac{(\gamma^k)^2}{\chi^k|\rho_2|}mC^2.
\end{aligned}
\end{equation}

Therefore, under Assumption \ref{ass:dp-noise} and the conditions for $\chi^k$ and $\gamma^k$ in (\ref{eq:condtions_chi}), we have that the sequence $\{\sum_{i=1}^{m}\|x_i^k- \bar{x}^k\|^2\}$ satisfies the conditions for $\{v^k\}$ in Lemma \ref{Lemma-polyak}, and hence, converges to zero almost surely. So
$x_i^k$ converges to $\bar{x}^k$ almost surely. Further recalling  $\bar{x}^k$ converging {\it a.s.} to $\bar{r}^k$ in Lemma \ref{le:bar_x=bar_r} yields that  $x_i^k$ converges {\it a.s.} to $\bar{r}^k$.

Moreover, Lemma \ref{Lemma-polyak} also implies  the following relation  {\it a.s.}:
\begin{equation}\label{eq:square}
 \sum_{k=0}^{\infty}\chi^k\sum_{i=1}^{m}\|x_i^k- \bar{x}^k\|^2 <\infty.
\end{equation}

To prove the last statement, we invoke the Cauchy–Schwarz inequality, which ensures
\[
\begin{aligned}
&\sum_{k=0}^{\infty} \sqrt{\chi^k\sum_{i=1}^{m}\|x_i^k- \bar{x}^k\|^2}\sqrt{\frac{(\gamma^k)^2}{\chi^k}}\\
&\qquad\qquad\leq \sqrt{\sum_{k=0}^{\infty} \chi^k\sum_{i=1}^{m}\|x_i^k- \bar{x}^k\|^2}\sqrt{\sum_{k=0}^{\infty} \frac{(\gamma^k)^2}{\chi^k}}.
\end{aligned}
\]
Noting that the summand in the left hand side of the preceding inequality is actually $\gamma^k\sqrt{\sum_{i=1}^{m}\|x_i^k- \bar{x}^k\|^2}$, and the right hand side of the preceding inequality is less than infinity almost surely under the proven result in (\ref{eq:square}) and the assumption in (\ref{eq:condtions_chi}), we have that
$\sum_{k=0}^{\infty}\gamma^k\sqrt{\sum_{i=1}^{m}\|x_i^k- \bar{x}^k\|^2}<\infty$ holds almost surely. Further utilizing the relationship
$\sum_{i=1}^{m}\|x_i^k- \bar{x}^k\|\leq  \sqrt{m\sum_{i=1}^{m}\|x_i^k- \bar{x}^k\|^2}$ yields the stated result.
\end{proof}

\begin{Remark 1}
In Theorem \ref{Th:consensus_tracking},  $\{\gamma^k\}$ decays to zero. In combination with Assumption \ref{as:reference_signal}, this requires $r^k$s' variations to  gradually decay with time, which is necessary to ensure every $x_i^k$ to track  the  {\it exact} average reference signal $\bar{r}^k$. This is because when individual reference  signals  are unknown and varying persistently, in general it is impossible for agents to precisely track their average using  discrete fixed-frequency samples  of these unknown  signals \cite{kia2019tutorial}. Note that this assumption is satisfied in many applications of dynamic average consensus. For example, in dynamic-average-consensus based distributed optimization, the reference signals of individual agents are gradients of individual objective functions, which gradually converge to constant values as the iterates converge to the optimal solution \cite{nedic2017achieving}. This is also the case in distributed Nash equilibrium seeking, where the reference signals of individual agents (usually called players in games) are individual pseudogradients, which will converge to constant values when the players converge to the Nash equilibrium \cite{koshal2016distributed}. 
\end{Remark 1}

\begin{Remark 1}
In distributed systems, usually  communication imperfections can be modeled as channel noises \cite{kar2008distributed}, which can be regarded as a special case of   DP-noise.  Therefore, Algorithm 1 can also be used to counteract  communication imperfections in distributed computation.
\end{Remark 1}

Next, we prove that Algorithm 1 can ensure $\epsilon$-DP for individual agents' reference signals $r_i^k$ with the cumulative privacy budget guaranteed to be finite, even when the number of iterations tends to infinity.

\section{Differential-privacy Analysis}

  We  have to characterize the sensitivity of Algorithm 1 in order to quantify the  level of  enabled privacy strength. Similar to  the sensitivity definition of iterative   optimization algorithms in \cite{Huang15}, we define  the sensitivity of a dynamic average consensus  algorithm as follows: 
\begin{Definition 1}\label{de:sensitivity}
  At each iteration $k$, for any initial state $\vartheta^0$ and any adjacent dynamic average consensus  problems  $\mathcal{P}$ and $\mathcal{P'}$,  the sensitivity of Algorithm 1 is
  \begin{equation}
  \Delta^k\triangleq \sup\limits_{\mathcal{O}\in\mathbb{O}}\left\{\sup\limits_{\vartheta\in\mathcal{R}_{\mathcal{P},\vartheta^0}^{-1}(\mathcal{O}),\:\vartheta'\in\mathcal{R}_{\mathcal{P}',\vartheta^0}^{-1}(\mathcal{O})}\hspace{-0.3cm}\|\vartheta^{k}-\vartheta'^{k}\|_1\right\}.
  \end{equation}
\end{Definition 1}

\begin{Lemma 1}\label{Le:Laplacian}
In Algorithm 1,  if each agent's DP-noise vector $\zeta_i^k\in\mathbb{R}^d$  consists of $d$ independent Laplace noises with  parameter $\nu^k$   such that $\sum_{k=1}^{T_0}\frac{\Delta^k}{\nu^k}\leq \bar\epsilon$, then Algorithm 1 is $\epsilon$-differentially private with the cumulative privacy budget from iteration   $k=0$ to $k=T_0$ less than $\bar\epsilon$.
\end{Lemma 1}
\begin{proof}
The lemma   follows the same line of reasoning of Lemma 2 in  \cite{Huang15} (also see Theorem 3 in \cite{ye2021differentially}).
\end{proof}

\begin{Theorem 1}\label{th:DP_Algorithm1}
Under  the conditions of Theorem \ref{Th:consensus_tracking}, if all elements of $\zeta_i^k$ are drawn independently from  Laplace distribution ${\rm Lap}(\nu^k)$ with $(\sigma_i^k)^2=2(\nu^k)^2$ satisfying Assumption \ref{ass:dp-noise},  then all agents
will converge {\it a.s.} to the average reference signal $\bar{r}^k$. Moreover,
\begin{enumerate}
\item For any finite number of iterations $T$, Algorithm 1 is  $\epsilon$-differentially private with the cumulative privacy budget bounded by $\epsilon\leq \sum_{k=1}^{T}\frac{2C_r\varsigma^k}{\nu^k}$   where $\varsigma^k\triangleq \sum_{p=1}^{k-1} \Pi_{q=p}^{k-1}(1-\alpha^q-\bar{L}\gamma^{q}) \Lambda^{p-1} +\Lambda^{k-1}$, $\bar{L}\triangleq\min_i\{|L_{ii}|\}$, $\Lambda^k \triangleq \gamma^{k+1}\chi^{k+1} +  (1-\alpha^k)\gamma^{k}\chi^{k}$, and $C_r$ is given in Definition \ref{de:adjacency};
\item  The cumulative privacy budget is  finite for $T\rightarrow\infty$  when the sequence  $\{\frac{\gamma^k}{\nu^k}\}$ is summable.
\end{enumerate}
\end{Theorem 1}
\begin{proof}
Since the Laplace noise satisfies Assumption \ref{ass:dp-noise}, the convergence result  follows directly from Theorem 1.

To prove the two statements on the strength of $\epsilon$-DP, we first analyze the sensitivity of the algorithm.
 Given two adjacent dynamic average consensus problems $\mathcal{P}$ and $\mathcal{P'}$, for any given fixed observation $\mathcal{O}$ and initial state $\vartheta^0= x^0$,   the sensitivity depends on $\|x^{k}-x'^{k}\|_1$ according to Definition \ref{de:sensitivity}. Since in $\mathcal{P}$ and $\mathcal{P'}$, only one reference signal  is different, we  represent this different reference signal as the  $i$th one, i.e., $r_i$ in $\mathcal{P}$ and $r'_i$ in $\mathcal{P}'$, without loss of generality.

Because the initial conditions, reference signals, and observations of $\mathcal{P}$ and $\mathcal{P'}$  are identical for $j\neq i$, we have $x_j^k={x'_j}^k$ for all $j\neq i$ and $k$. Therefore, $\|x^{k}-x'^{k}\|_1$ is always equal to $\|x_i^{k}-{x'_i}^{k}\|_1$.

   Algorithm 1   implies
 \[
 \begin{aligned}
x_i^{k+1}-{x'_i}^{k+1}=&(1-\alpha^k-|L_{ii}|\chi^k)(x_i^k-{x'_i}^k)\\
 &+ (r_i^{k+1}-{r'_i}^{k+1})-(1-\alpha^k) (r_i^k-{r'_i}^k).
 \end{aligned}
 \]
 Note that we have  used the  fact that the observations $x_j^k+\zeta_j^k$ and ${x'_j}^k+{\zeta'_j}^k$ are the same.

 Hence, using the third condition in Definition \ref{de:adjacency}, the sensitivity $\Delta^k$ satisfies
 \begin{equation}\label{eq:sensitivity_iterationi}
 \begin{aligned}
 &\Delta^{k+1}\\
 &\leq (1-\alpha^k-|L_{ii}|\chi^k)\Delta^{k}+ C_r\gamma^{k+1}\chi^{k+1} + C_r(1-\alpha^k)\gamma^{k}\chi^{k}\\
 &\leq (1-\alpha^k-\bar{L}\chi^k)\Delta^{k}+ C_r\Lambda^k,
 \end{aligned}
 \end{equation}
 where  $\bar{L}$ and $\Lambda^k$ are defined in the theorem statement. Then, by iteration, we can arrive at the first privacy statement   using Lemma \ref{Le:Laplacian} (note $\Delta^0=0$ as $\mathcal{P}$ and $\mathcal{P}'$ have identical initial conditions).

 For the infinity-time-horizon result in the second statement, we exploit Lemma \ref{le:chung} and the third condition in Definition \ref{de:adjacency}.  More specifically, from (\ref{eq:sensitivity_iterationi}), according to Lemma \ref{le:chung},  we can always find some $\bar{C}$ such that $\Delta^k\leq  \bar{C} \gamma^k $ holds (note that $\alpha^k$ decays faster than $\chi^k$).
Using Lemma \ref{Le:Laplacian}, we can easily obtain $\epsilon\leq \sum_{k=1}^{T}\frac{\bar{C}\gamma^k}{ \nu^k}$. Hence, $\epsilon$ will  be finite even when $T$ tends to infinity if  the sequence $\{\frac{\gamma^k}{\nu^k}\}$ is summable.
\end{proof}

Note that in dynamic average consensus applications such as Nash equilibrium seeking, \cite{ye2021differentially} achieves  $\epsilon$-DP by enforcing the tracking input signal  to be summable (by multiplying it with a geometrically-decreasing factor), which, however, also makes it impossible to ensure accurate convergence to the  desired equilibrium. In our approach, by allowing the tracking input $r_i^{k+1}-(1-\alpha^k)r_i^k$  to be non-summable (since we allow $\sum_{k=0}^{\infty}\gamma^k=\infty$), we achieve both accurate convergence and finite cumulative privacy budget, even when the number of iterations goes to infinity.  To our knowledge, this is the first dynamic average consensus  algorithm that can achieve both almost sure convergence to the {\it exact} average reference signal and rigorous $\epsilon$-DP, even with the number of iterations going to infinity.

\begin{Remark 1}
  To ensure that the cumulative privacy budget $ \sum_{k=1}^{\infty}\frac{\bar{C}\gamma^k}{\nu^k} $ is bounded, we  employ   Laplace noise with parameter $\nu^k$  increasing with time (since we require the sequence $\{\frac{\gamma^k}{\nu^k}\}$ to be summable while the sequence $\{ \gamma^k \}$ is non-summable).  Because the strength of the shared signal is time-invariant (i.e., $x_i^k$), an increasing $\nu^k$  makes the relative level between noise $\zeta_i^k$ and signal $x_i^k$ increase  with time. However, since it is $\chi^k{\rm Lap}(\nu^k)$ that is actually  fed into the algorithm, and  the increase  in the noise level  $\nu^k$ is outweighed by the decrease of $\chi^k$ (see  Assumption \ref{ass:dp-noise}), the actual noise fed into the algorithm  still decays with time, which explains why  Algorithm 1 can ensure every agent's  accurate convergence. Moreover, according to Theorem \ref{Th:consensus_tracking}, the convergence will not be affected if we  scale  $\nu^k$ by any constant  $\frac{1}{\epsilon}>0$  to achieve any desired level of $\epsilon$-DP, as long as  $\nu^k$ (with associated variance $(\sigma_i^k)^2=2(\nu^k)^2$) satisfies Assumption \ref{ass:dp-noise}. 
\end{Remark 1}

\begin{Remark 1}
 It is worth noting that   our   simultaneous achievement of both provable accurate convergence and $\epsilon$-DP does not contradict the fundamental  limitations of the DP theory \cite{dwork2014algorithmic}. In fact,  the DP theory indicates that a query mechanism  on a dataset can achieve  $\epsilon$-DP only by sacrificing the accuracy of query. However, in dynamic average consensus, what are queried in every iteration  are individual reference signals but not the average reference signal, and  revealing the value of the average reference signal at steady state to an observer  is not equivalent to revealing entire individual reference signals (the actual query target).
\end{Remark 1}


\section{Numerical Simulations}\label{se:simulation}
\begin{figure}
\includegraphics[width=0.5\textwidth]{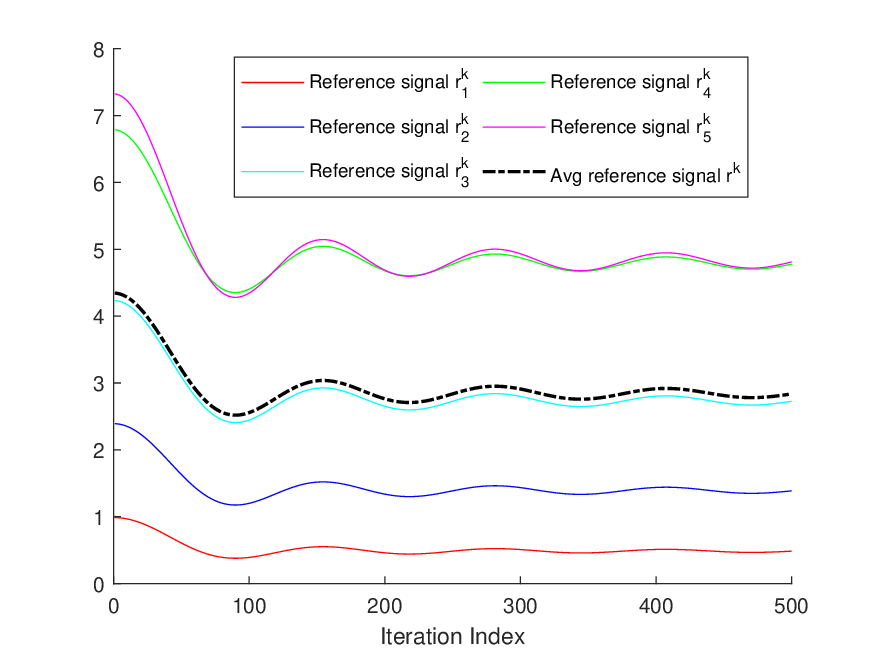}
    \caption{The individual reference signals of five agents and the average reference signal (dashed black curve).}
    \label{fig:signals}
    \vspace{-0.3cm}
\end{figure}

\begin{figure}
\includegraphics[width=0.5\textwidth]{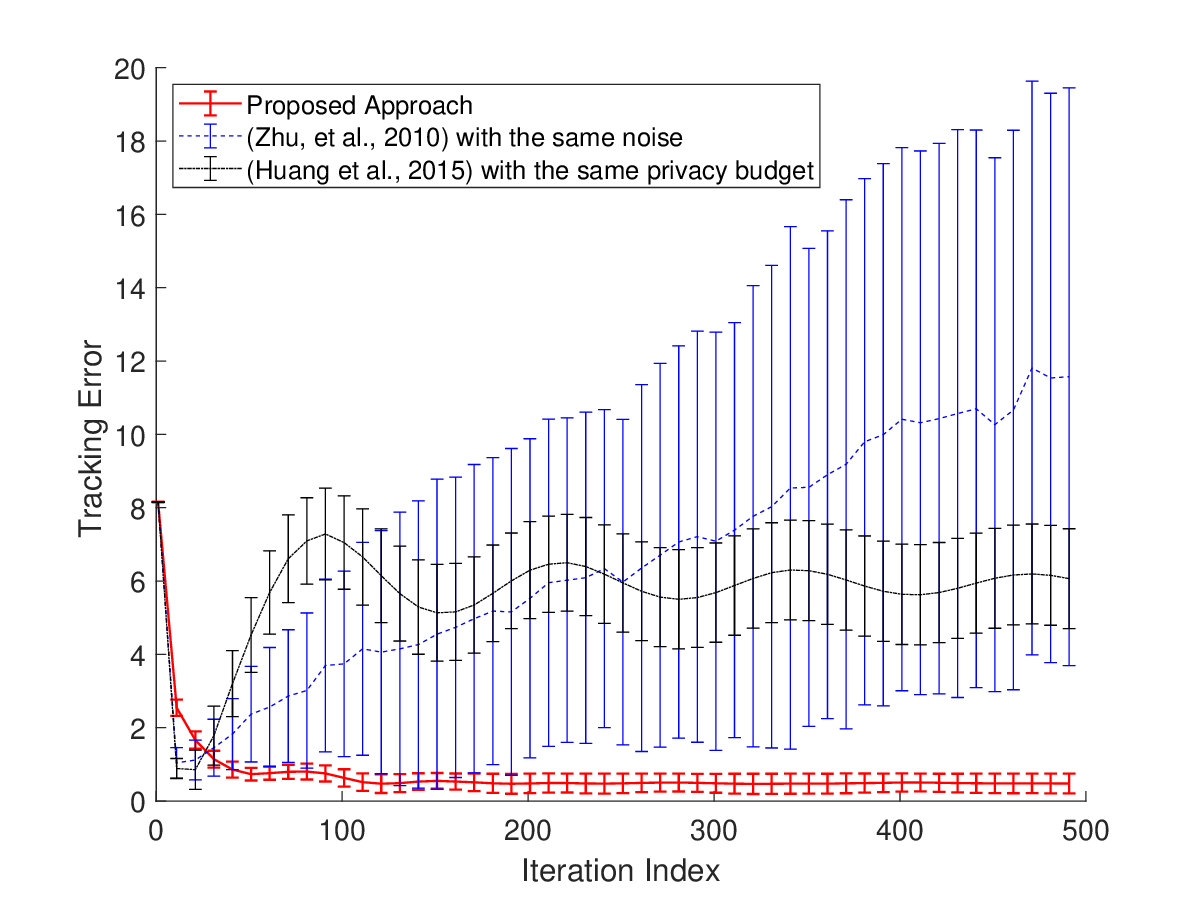}
    \caption{Comparison of the proposed algorithm  with existing DP approach for distributed optimization from Huang et al. 2015 \cite{Huang15} which can be adapted to dynamic average consensus (under the same privacy budget $\epsilon$)  and the conventional dynamic average consensus approach in Zhu et al., 2010 \cite{zhu2010discrete} (under the same noise). The tracking error is defined as ($\sum_{i=1}^{m}\|x_i^k-\bar{x}^k\|$). }
    \label{fig:tracking_error}
\end{figure}

We evaluate the performance of the  proposed  algorithm using a network of $m=5$ agents. In the simulation, we set the reference signal of agent $i$ as $a_i+\frac{b_i}{10k}\sin(0.05k)$ where $a_i$ and $b_i$ are randomly selected from a uniform distribution on $(0,\,10)$ for all $i\in[m]$. The used reference signals for all agents as well as the average reference signal are depicted in Fig. \ref{fig:signals}.  To enable DP, we inject Laplace noise with parameter $\nu^k=1+0.1k^{0.2}$.  We set the  diminishing sequence  as   $\chi^k =\frac{2}{1+ k^{0.9}}$. The stepsize is set as $\alpha^k=\frac{0.01}{1+k}$. It can be verified that  the conditions in Theorem 1 are satisfied under the setting. In the evaluation, we run our algorithm for 100 times, and calculate the average and the variance of the tracking error $\sum_{i=1}^{m}\|x_i^k-\bar{x}^k\|$ against the iteration index $k$. The result is given by the red curve and error bars in Fig. \ref{fig:tracking_error}. For comparison, we also run the existing dynamic average consensus algorithm  proposed by Zhu et al. in \cite{zhu2010discrete} under the same noise, and the existing DP approach   proposed by Huang et al. in \cite{Huang15} under the same cumulative privacy budget $\epsilon$. Note that the DP approach in \cite{Huang15} addresses  distributed optimization problems, which can be viewed as a dynamic average consensus problem if we regard the gradient of an agent's objective function  as the  input to the agent. We adapt  its DP mechanism (geometrically decreasing stepsizes for input gradients  and geometrically decreasing DP-noise) to the dynamic average consensus problem.   The evolution of the average error/variance of the  approaches in \cite{zhu2010discrete} and \cite{Huang15}  are given by  the blue and black curves/error bars in Fig. \ref{fig:tracking_error}, respectively. It can be seen that the proposed algorithm has a   much better  accuracy.

\begin{figure}
\center
    \vspace{-0.7cm}
    \includegraphics[width=0.25\textwidth]{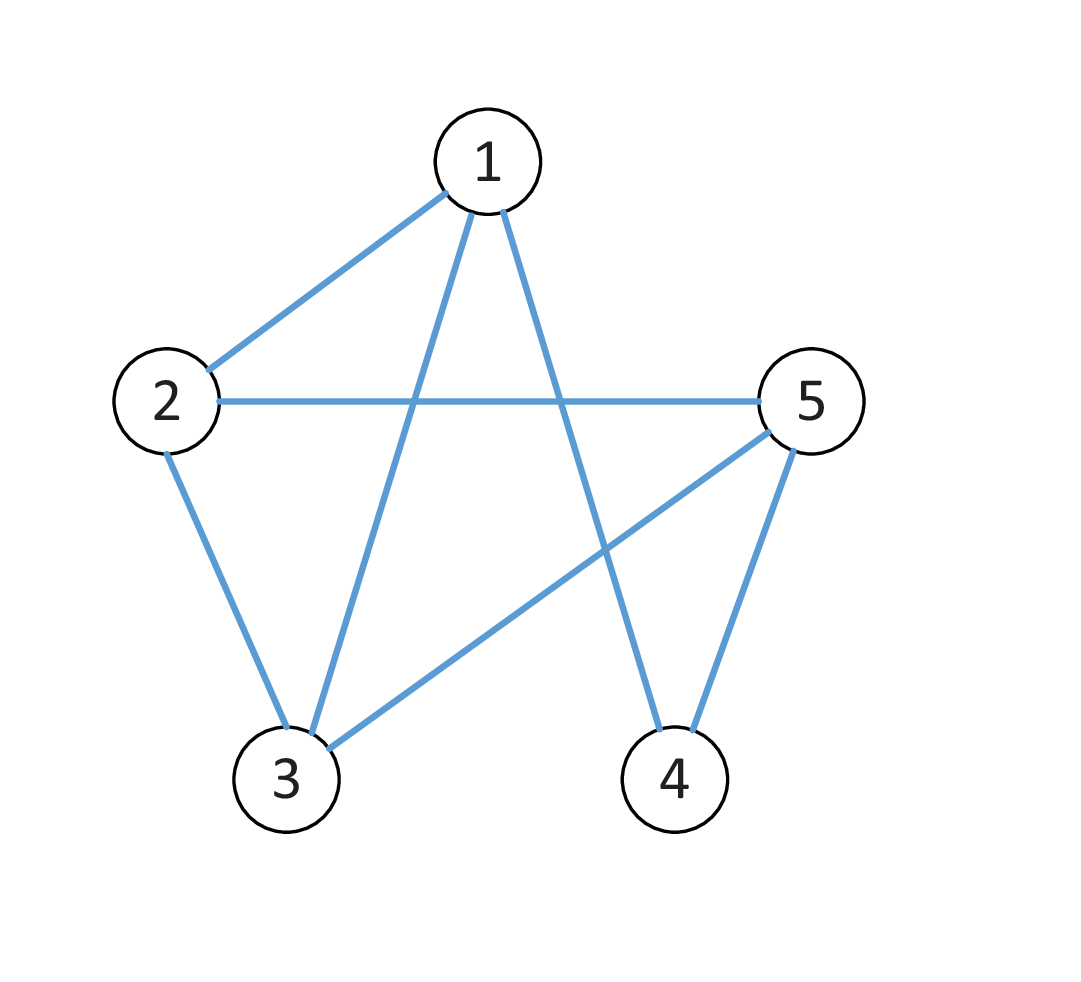}
    \caption{The interaction network used in numerical simulation.}
    \label{fig:network}
    \vspace{-.2cm}
\end{figure}

\section{Conclusions}\label{se:conclusions}

This paper proposes a robust dynamic average consensus algorithm that can ensure provable convergence to the exact average reference signal even in the presence of persistent differential-privacy noise. Different from  existing differential privacy solutions for average consensus that have to trade convergence accuracy for differential privacy,  the proposed approach can    ensure both provable convergence to the exact desired average signal and rigorous $\epsilon$-differential privacy. To our knowledge, the simultaneous achievement of both $\epsilon$-differential privacy and provable convergence accuracy has not been  achieved in average consensus before.  Since channel noise in communication can  be viewed  as a special case of differential-privacy noise, our algorithm can also counteract communication imperfections in dynamic average consensus.   Numerical simulation   results   confirm the effectiveness of the proposed algorithm.


\bibliographystyle{IEEEtran}

\bibliography{reference1}

\begin{thebibliography}{10}
\providecommand{\url}[1]{#1}
\csname url@samestyle\endcsname
\providecommand{\newblock}{\relax}
\providecommand{\bibinfo}[2]{#2}
\providecommand{\BIBentrySTDinterwordspacing}{\spaceskip=0pt\relax}
\providecommand{\BIBentryALTinterwordstretchfactor}{4}
\providecommand{\BIBentryALTinterwordspacing}{\spaceskip=\fontdimen2\font plus
\BIBentryALTinterwordstretchfactor\fontdimen3\font minus
  \fontdimen4\font\relax}
\providecommand{\BIBforeignlanguage}[2]{{%
\expandafter\ifx\csname l@#1\endcsname\relax
\typeout{** WARNING: IEEEtran.bst: No hyphenation pattern has been}%
\typeout{** loaded for the language `#1'. Using the pattern for}%
\typeout{** the default language instead.}%
\else
\language=\csname l@#1\endcsname
\fi
#2}}
\providecommand{\BIBdecl}{\relax}
\BIBdecl

\bibitem{wang2022ensure}
Y.~Wang, ``Ensure differential privacy and convergence accuracy in consensus
  tracking and aggregative games with coupling constraints,'' \emph{arXiv
  preprint arXiv:2210.16395}, 2022.

\bibitem{zhu2010discrete}
M.~Zhu and S.~Mart{\'\i}nez, ``Discrete-time dynamic average consensus,''
  \emph{Automatica}, vol.~46, no.~2, pp. 322--329, 2010.

\bibitem{kia2019tutorial}
S.~S. Kia, B.~Van~Scoy, J.~Cortes, R.~A. Freeman, K.~M. Lynch, and S.~Martinez,
  ``Tutorial on dynamic average consensus: The problem, its applications, and
  the algorithms,'' \emph{IEEE Control Systems Magazine}, vol.~39, no.~3, pp.
  40--72, 2019.

\bibitem{yang2007distributed}
P.~Yang, R.~A. Freeman, and K.~M. Lynch, ``Distributed cooperative active
  sensing using consensus filters,'' in \emph{Proceedings 2007 IEEE
  International Conference on Robotics and Automation}.\hskip 1em plus 0.5em
  minus 0.4em\relax IEEE, 2007, pp. 405--410.

\bibitem{olfati2005consensus}
R.~Olfati-Saber and J.~S. Shamma, ``Consensus filters for sensor networks and
  distributed sensor fusion,'' in \emph{Proceedings of the 44th IEEE Conference
  on Decision and Control}.\hskip 1em plus 0.5em minus 0.4em\relax IEEE, 2005,
  pp. 6698--6703.

\bibitem{george2019robust}
J.~George and R.~A. Freeman, ``Robust dynamic average consensus algorithms,''
  \emph{IEEE Transactions on Automatic Control}, vol.~64, no.~11, pp.
  4615--4622, 2019.

\bibitem{koshal2016distributed}
J.~Koshal, A.~Nedi{\'c}, and U.~V. Shanbhag, ``Distributed algorithms for
  aggregative games on graphs,'' \emph{Operations Research}, vol.~64, no.~3,
  pp. 680--704, 2016.

\bibitem{belgioioso2020distributed}
G.~Belgioioso, A.~Nedi{\'c}, and S.~Grammatico, ``Distributed generalized
  {Nash} equilibrium seeking in aggregative games on time-varying networks,''
  \emph{IEEE Transactions on Automatic Control}, vol.~66, no.~5, pp.
  2061--2075, 2021.

\bibitem{xu2015augmented}
J.~Xu, S.~Zhu, Y.~C. Soh, and L.~Xie, ``Augmented distributed gradient methods
  for multi-agent optimization under uncoordinated constant stepsizes,'' in
  \emph{Proceedings of IEEE Conference on Decision and Control}.\hskip 1em plus
  0.5em minus 0.4em\relax IEEE, 2015, pp. 2055--2060.

\bibitem{di2016next}
P.~Di~Lorenzo and G.~Scutari, ``{NEXT}: In-network nonconvex optimization,''
  \emph{IEEE Transactions on Signal and Information Processing over Networks},
  vol.~2, no.~2, pp. 120--136, 2016.

\bibitem{pu2020push}
S.~Pu, W.~Shi, J.~Xu, and A.~Nedi\'{c}, ``Push-pull gradient methods for
  distributed optimization in networks,'' \emph{IEEE Transactions on Automatic
  Control}, vol.~66, no.~1, pp. 1--16, 2021.

\bibitem{zhang2019admm}
C.~Zhang, M.~Ahmad, and Y.~Wang, ``{ADMM} based privacy-preserving
  decentralized optimization,'' \emph{IEEE Transactions on Information
  Forensics and Security}, vol.~14, no.~3, pp. 565--580, 2019.

\bibitem{burbano2019inferring}
D.~A. Burbano-L, J.~George, R.~A. Freeman, and K.~M. Lynch, ``Inferring private
  information in wireless sensor networks,'' in \emph{IEEE International
  Conference on Acoustics, Speech and Signal Processing (ICASSP)}.\hskip 1em
  plus 0.5em minus 0.4em\relax IEEE, 2019, pp. 4310--4314.

\bibitem{fang2011smart}
X.~Fang, S.~Misra, G.~Xue, and D.~Yang, ``Smart grid—the new and improved
  power grid: A survey,'' \emph{IEEE Communications Surveys \& Tutorials},
  vol.~14, no.~4, pp. 944--980, 2011.

\bibitem{yan2012distributed}
F.~Yan, S.~Sundaram, S.~Vishwanathan, and Y.~Qi, ``Distributed autonomous
  online learning: Regrets and intrinsic privacy-preserving properties,''
  \emph{IEEE Transactions on Knowledge and Data Engineering}, vol.~25, no.~11,
  pp. 2483--2493, 2012.

\bibitem{wang2022tailoring}
Y.~Wang and A.~Nedi{\'c}, ``Tailoring gradient methods for
  differentially-private distributed optimization,'' \emph{arXiv preprint
  arXiv:2202.01113}, 2022.

\bibitem{wang2022decentralized}
Y.~Wang and H.~V. Poor, ``Decentralized stochastic optimization with inherent
  privacy protection,'' \emph{IEEE Transactions on Automatic Control}, 2022.

\bibitem{zhu2019deep}
L.~Zhu, Z.~Liu, and S.~Han, ``Deep leakage from gradients,'' \emph{Advances in
  Neural Information Processing Systems}, vol.~32, 2019.

\bibitem{kia2015dynamic}
S.~S. Kia, J.~Cort{\'e}s, and S.~Martinez, ``Dynamic average consensus under
  limited control authority and privacy requirements,'' \emph{International
  Journal of Robust and Nonlinear Control}, vol.~25, no.~13, pp. 1941--1966,
  2015.

\bibitem{zhang2022privacy}
K.~Zhang, Z.~Li, Y.~Wang, A.~Louati, and J.~Chen, ``Privacy-preserving dynamic
  average consensus via state decomposition: Case study on multi-robot
  formation control,'' \emph{Automatica}, vol. 139, p. 110182, 2022.

\bibitem{gao2018privacy}
H.~Gao, C.~Zhang, M.~Ahmad, and Y.~Wang, ``Privacy-preserving average consensus
  on directed graphs using push-sum,'' in \emph{IEEE Conference on
  Communications and Network Security (CNS)}.\hskip 1em plus 0.5em minus
  0.4em\relax IEEE, 2018, pp. 1--9.

\bibitem{ruan2019secure}
M.~Ruan, H.~Gao, and Y.~Wang, ``Secure and privacy-preserving consensus,''
  \emph{IEEE Transactions on Automatic Control}, vol.~64, no.~10, pp.
  4035--4049, 2019.

\bibitem{wang2019privacy}
Y.~Wang, ``Privacy-preserving average consensus via state decomposition,''
  \emph{IEEE Transactions on Automatic Control}, vol.~64, no.~11, pp.
  4711--4716, 2019.

\bibitem{gao2022algorithm}
H.~Gao and Y.~Wang, ``Algorithm-level confidentiality for average consensus on
  time-varying directed graphs,'' \emph{IEEE Transactions on Network Science
  and Engineering}, vol.~9, no.~2, pp. 918--931, 2022.

\bibitem{manitara2013privacy}
N.~E. Manitara and C.~N. Hadjicostis, ``Privacy-preserving asymptotic average
  consensus,'' in \emph{2013 European Control Conference (ECC)}.\hskip 1em plus
  0.5em minus 0.4em\relax IEEE, 2013, pp. 760--765.

\bibitem{pequito2014design}
S.~Pequito, S.~Kar, S.~Sundaram, and A.~P. Aguiar, ``Design of communication
  networks for distributed computation with privacy guarantees,'' in \emph{53rd
  IEEE Conference on Decision and Control}.\hskip 1em plus 0.5em minus
  0.4em\relax IEEE, 2014, pp. 1370--1376.

\bibitem{mo2016privacy}
Y.~Mo and R.~M. Murray, ``Privacy preserving average consensus,'' \emph{IEEE
  Transactions on Automatic Control}, vol.~62, no.~2, pp. 753--765, 2016.

\bibitem{nozari2017differentially}
E.~Nozari, P.~Tallapragada, and J.~Cort{\'e}s, ``Differentially private average
  consensus: Obstructions, trade-offs, and optimal algorithm design,''
  \emph{Automatica}, vol.~81, pp. 221--231, 2017.

\bibitem{gupta2017privacy}
N.~Gupta, J.~Katz, and N.~Chopra, ``Privacy in distributed average consensus,''
  \emph{IFAC-PapersOnLine}, vol.~50, no.~1, pp. 9515--9520, 2017.

\bibitem{altafini2019dynamical}
C.~Altafini, ``A dynamical approach to privacy preserving average consensus,''
  in \emph{2019 IEEE 58th Conference on Decision and Control (CDC)}.\hskip 1em
  plus 0.5em minus 0.4em\relax IEEE, 2019, pp. 4501--4506.

\bibitem{he2018privacy}
J.~He, L.~Cai, C.~Zhao, P.~Cheng, and X.~Guan, ``Privacy-preserving average
  consensus: privacy analysis and algorithm design,'' \emph{IEEE Transactions
  on Signal and Information Processing over Networks}, vol.~5, no.~1, pp.
  127--138, 2018.

\bibitem{dwork2014algorithmic}
C.~Dwork, A.~Roth \emph{et~al.}, ``The algorithmic foundations of differential
  privacy.'' \emph{Foundations and Trends in Theoretical Computer Science},
  vol.~9, no. 3-4, pp. 211--407, 2014.

\bibitem{huang2012differentially}
Z.~Huang, S.~Mitra, and G.~Dullerud, ``Differentially private iterative
  synchronous consensus,'' in \emph{Proceedings of the 2012 ACM workshop on
  Privacy in the Electronic Society}.\hskip 1em plus 0.5em minus 0.4em\relax
  ACM, 2012, pp. 81--90.

\bibitem{fiore2019resilient}
D.~Fiore and G.~Russo, ``Resilient consensus for multi-agent systems subject to
  differential privacy requirements,'' \emph{Automatica}, vol. 106, pp. 18--26,
  2019.

\bibitem{he2020differential}
J.~He, L.~Cai, and X.~Guan, ``Differential private noise adding mechanism and
  its application on consensus algorithm,'' \emph{IEEE Transactions on Signal
  Processing}, vol.~68, pp. 4069--4082, 2020.

\bibitem{liu2020differentially}
X.-K. Liu, J.-F. Zhang, and J.~Wang, ``Differentially private consensus
  algorithm for continuous-time heterogeneous multi-agent systems,''
  \emph{Automatica}, vol. 122, p. 109283, 2020.

\bibitem{dwork2010differential}
C.~Dwork, M.~Naor, T.~Pitassi, and G.~N. Rothblum, ``Differential privacy under
  continual observation,'' in \emph{Proceedings of the forty-second ACM
  Symposium on Theory of Computing}, 2010, pp. 715--724.

\bibitem{Huang15}
Z.~Huang, S.~Mitra, and N.~Vaidya, ``Differentially private distributed
  optimization,'' in \emph{Proceedings of the 2015 International Conference on
  Distributed Computing and Networking}, New York, NY, USA, 2015.

\bibitem{kairouz2015composition}
P.~Kairouz, S.~Oh, and P.~Viswanath, ``The composition theorem for differential
  privacy,'' in \emph{International Conference on Machine Learning}.\hskip 1em
  plus 0.5em minus 0.4em\relax PMLR, 2015, pp. 1376--1385.

\bibitem{bun2016concentrated}
M.~Bun and T.~Steinke, ``Concentrated differential privacy: Simplifications,
  extensions, and lower bounds,'' in \emph{Theory of Cryptography
  Conference}.\hskip 1em plus 0.5em minus 0.4em\relax Springer, 2016, pp.
  635--658.

\bibitem{mironov2017renyi}
I.~Mironov, ``R{\'e}nyi differential privacy,'' in \emph{The 30th Computer
  Security Foundations Symposium}.\hskip 1em plus 0.5em minus 0.4em\relax IEEE,
  2017, pp. 263--275.

\bibitem{pu2020robust}
S.~Pu, ``A robust gradient tracking method for distributed optimization over
  directed networks,'' in \emph{IEEE Conference on Decision and Control
  (CDC)}.\hskip 1em plus 0.5em minus 0.4em\relax IEEE, 2020, pp. 2335--2341.

\bibitem{wang2022gradient}
Y.~Wang and T.~Ba{\c{s}}ar, ``Gradient-tracking based distributed optimization
  with guaranteed optimality under noisy information sharing,'' \emph{IEEE
  Transactions on Automatic Control}, 2022.

\bibitem{nedic2017achieving}
A.~Nedic, A.~Olshevsky, and W.~Shi, ``Achieving geometric convergence for
  distributed optimization over time-varying graphs,'' \emph{SIAM Journal on
  Optimization}, vol.~27, no.~4, pp. 2597--2633, 2017.

\bibitem{kar2008distributed}
S.~Kar and J.~M. Moura, ``Distributed consensus algorithms in sensor networks
  with imperfect communication: Link failures and channel noise,'' \emph{IEEE
  Transactions on Signal Processing}, vol.~57, no.~1, pp. 355--369, 2008.

\bibitem{ye2021differentially}
M.~Ye, G.~Hu, L.~Xie, and S.~Xu, ``Differentially private distributed {Nash}
  equilibrium seeking for aggregative games,'' \emph{IEEE Transactions on
  Automatic Control}, vol.~67, no.~5, pp. 2451--2458, 2021.

\end{thebibliography}

\end{document}